\documentclass[conference]{IEEEtran}

\usepackage{graphicx}

\makeatletter
\let\MYcaption\@makecaption
\makeatother

\usepackage{subcaption}

\makeatletter
\let\@makecaption\MYcaption
\makeatother

\usepackage{amsmath}
\usepackage{amssymb}

\usepackage{booktabs}
\usepackage{siunitx}

\usepackage{todonotes}

\usepackage{tikz}
\usetikzlibrary{shapes}
\usepackage{pgfplots}
\pgfplotsset{compat=1.14}
\definecolor{color0}{rgb}{0.12156862745098,0.466666666666667,0.705882352941177}
\definecolor{color1}{rgb}{1,0.498039215686275,0.0549019607843137}
\definecolor{color2}{rgb}{0.172549019607843,0.627450980392157,0.172549019607843}

\usepackage{cite}

\begin{document}

\title{Power Inversion of the Massive MIMO Channel}
\author{
\IEEEauthorblockN{Jens Abraham \hspace{4.8cm} Torbj\"orn Ekman}
\IEEEauthorblockA{
    \textit{Circuit and Radio Systems Group, Department of Electronic Systems} \\
    \textit{Norwegian University of Science and Technology} \\
    Trondheim, Norway \\
    \texttt{\{jens.abraham,torbjorn.ekman\}@ntnu.no}
}
}

\maketitle

\begin{abstract}
Channel hardening characterises the diminishing influence of small scale fading on large scale antenna systems.
The effective massive MIMO time domain channel is introduced and applied to a maximum diversity channel with rectangular power delay profile.
This model bounds channel hardening and allows a proper interpretation from a radio design perspective.
The reduced variability of the effective channel enables power inversion to obtain a downlink channel that only depends on the large scale fading properties.
\end{abstract}
\begin{IEEEkeywords}
massive MIMO, time reversal, channel hardening, power inversion
\end{IEEEkeywords}

\section{Introduction}
Wireless sensor networks (WSN) are of increasing interest to industry and governments for surveillance of different environments.
Large scale antenna systems (such as massive MIMO base stations (BS) \cite{marzetta_massive_2015}) could become a leading technology to provide a robust single hop data link for thousands of sensor nodes.
They make it possible to move complexity from the sensors to the BS to increase the lifetime of each node.
Channel hardening and favourable propagation allow the simplification of the node transceiver design and a reduction of their output power.
An alternative approach are mesh networks, but they suffer from uneven power usage for nodes close to the data gateway.

Even though a single radio channel can experience small and large scale fading (here pathloss and shadowing), it is highly unlikely that all antenna elements experience a fading dip at the same time.
Thus, large scale antenna systems can exploit spatial diversity to compensate for small scale fading.
Furthermore, the array gain can overcome some large scale fading.

This paper formulates an effective massive MIMO channel in the time domain, to describe the small scale fading in the downlink with a relative power measure.
A similar approach in the frequency domain was chosen by the authors to investigate the behaviour of rms delay spread under channel hardening \cite{ghiaasi_measured_2018}.
The effective channel can directly be used to bound the fading margin and adheres to the philosophy that a receiver requires first and foremost a signal level above or at a minimal threshold.
Both centralised and distributed normalisations of time reversal precoding and their influence on the remaining small scale fading are studied.
In addition, focus is placed on the relative antenna element and BS power.
The former is defining the required dynamic range of the BS transmitters.
The latter is mainly of regulatory interest, but confines the overall power consumption of the BS in addition.

The next section describes the effective massive MIMO channel with consideration of time reversal precoding, normalisation and relative power measures.
The following section shows the distributions for the relative power measures of a maximum diversity channel to give a best case bound of channel hardening.
Afterwards, the results are applied to a four tap channel to demonstrate the ideal theoretical behaviour of large scale antenna systems with a growing number of transmitters.
The last section summarises the findings and discusses necessary steps to realise time reversal power inversion for robust large scale antenna system WSNs.

\section{The Effective Massive MIMO Channel}
The complex valued input-output relation at time index $n$ for downlink signal $x_l[n]$ intended for user $l$ and signal $y_k[n]$ received by user $k$ in a $K$ user system with $M$ antennas at the BS is described by
\begin{align}
    y_k[n] &= \sqrt{\beta_k} \sum_{l=1}^K \underbrace{\left( \sum_{m=1}^M h_{mk}[n] \star w_{ml}[n] \right)}_{\mathfrak{h}_{kl}[n]} \star x_l[n] + e_k[n] \notag \\
    &= \underbrace{\sqrt{\beta_k} \mathfrak{h}_{kk}[n] \star x_k[n]}_\text{signal} + \underbrace{\sqrt{\beta_k} \sum_{\substack{l=1\\l \neq k}}^K  \mathfrak{h}_{kl}[n] \star x_l[n]}_{\text{multi-user interference}} + \underbrace{e_k[n]}_{\text{noise}} \label{eqn:eff_channel}
\end{align}
where $\beta_k$, $e_k$ are large scale fading coefficient and noise, $h_{mk}[n]$ and $w_{ml}[n]$ are small scale fading channel impulse response for user $k$ and precoding filter for user $l$ transmitted from antenna $m$. The $\star$ denotes the convolution between two signals.
Here $\beta_k$ normalises the channel impulse response as
\begin{equation}
    \mathcal{E}\left\{\sum_{n=1}^N |h_{mk}[n]|^2 \right\} = 1 \label{eqn:pdp_normalisation}
\end{equation}
with $\mathcal{E}\{(.)\}$ denoting the expectation.
Intrinsically, $\beta_k$ is a global variable for all SISO channels from a user to the $M$ BS antennas.

The effective channels $\mathfrak{h}_{kl}[n]$ are formed by the superposition of all signals from the BS at the user $k$.
The intended effective channel is $\mathfrak{h}_{kk}[n]$, whereas all other effective channels contribute to multi-user interference.

\subsection{Time Reversal}
Maximum ratio transmission (MRT), zero forcing (ZF) and linear minimum mean square error (MMSE) \cite{lo_maximum_1999,joham_linear_2005} are the commonly used frequency domain linear precoding schemes in large scale antenna systems. 
MRT optimises the signal to noise ratio of a single user, ignoring multi-user interference (MUI).
ZF optimises the signal to interference ratio by supressing MUI, ignoring the SNR of the intended user and the linear MMSE precoder has a control parameter to achieve a trade off between MRT and ZF.

Both ZF and MMSE require a matrix inversion operation of the multi user channel to calculate the precoding weights.
The matrix inversion introduces the requirement of centralised weight calculations and is a computational heavy operation.
It is our understanding that favourable propagation and user scheduling can alleviate the MUI in a heavily loaded large scale antenna system for WSN.
Therefore, MRT will be the inspiration for the considered precoding scheme in the remainder of this paper.

MRT is usually applied to each sub-carrier of an orthogonal frequency-division multiplexing (OFDM) system and closely related to time reversal (TR) \cite{rusek_scaling_2013}.
Following the TR idea, the precoder weights can be calculated from the uplink channel with generic single user normalisation $c_l$
\begin{equation}
    w_{ml}[n] = \frac{h_{ml}^\ast[-n]}{c_l} \label{eqn:time_reversal_weights}
\end{equation}
where $^\ast$ denotes the complex conjugate.
This approach reverses the channel impulse response to focus energy at the user in both space and time \cite{oestges_characterization_2005}, partly reducing interference at other places.
The importance of the effective zero delay tap $\mathfrak{h}_{kk}[0]$ becomes apparent by investigation of the sum over convolutions in Eqn. \eqref{eqn:eff_channel}.
It is the main contributor to the effective channel due to the coherent addition of the underlying SISO channel taps.
Solving the convolutions for zero delay results in
\begin{equation}
    \mathfrak{h}_{kk}[0] = \frac{1}{c_k} \sum_{m=1}^M \sum_{n=1}^N |h_{mk}[n]|^2. \label{eqn:zero_delay_tap}
\end{equation}
This result describes the radio propagation between the BS and the user in a compressed form and captures the usable signal power for a single tap receiver.
The remaining variability of $\mathfrak{h}_{kk}[0]$ is due to the uncompensated small scale fading.

\subsection{Powers}
At the BS, both the \emph{relative antenna element transmit power} $P_{mk}^\text{Ant,R}$ and the \emph{relative BS transmit power} $P_{k}^\text{BS,R}$ are random variables of interest.
The former is describing how much the output power of each antenna is influenced by the precoding weights:
\begin{equation}
    P_{mk}^\text{Ant,R} = M \sum_{n=1}^N \left|w_{mk}[n]\right|^2. \label{eqn:p_ant}
\end{equation}
The distribution of $P_{mk}^\text{Ant,R}$ characterises how much each transmitter at the BS has to cope with fluctuations of the antenna element output power.
Furthermore, $P_{k}^\text{BS,R}$ sums over all squared weights of a specific user to see the impact on the whole BS:
\begin{equation}
    P_{k}^\text{BS,R} = \sum_{m=1}^M \sum_{n=1}^N \left|w_{mk}[n]\right|^2. \label{eqn:p_bs}
\end{equation}

At the user, the \emph{relative effective received power} captures the array gain normalised power fluctuation for a single tap receiver
\begin{equation}
    P_k^\text{RX,R} = \frac{1}{M} \left|\mathfrak{h}_{kk}[0]\right|^2. \label{eqn:p_rx}
\end{equation}
These fluctuations describe the remaining small scale fading and hereby how much spatial diversity is exploited by the precoding.

\subsection{Normalisations}
The three relative powers of interest are influenced by the choice of $c_l$ in Eqn. \eqref{eqn:time_reversal_weights}.
To implement a normalisation reminiscent to MRT, the normalisation constant has to be calculated by
\begin{equation}
    c_l^\text{TR} = \sqrt{\sum_{m=1}^M \sum_{n=1}^N |h_{ml}[n]|^2}. \label{eqn:tr_normalisation}
\end{equation}
This scales each realisation of the precoding weights with the current state of the channel and ensures unit gain per user over the whole BS.
Unfortunately, it requires a centralised weight calculation.
However, the double sum can be replaced by its expectation, leading to a decentralised strategy.
The inner sum follows Eqn. \eqref{eqn:pdp_normalisation} with expectation one and the outer sum is self-averaging over values fluctuating around one.
Eventually, a distributed time reversal (DTR) normalisation can be obtained as:
\begin{equation}
    c_l^\text{DTR} = \sqrt{\mathcal{E}\left\{\sum_{m=1}^M \sum_{n=1}^N |h_{ml}[n]|^2 \right\}} = \sqrt{M}.
\end{equation}

A third normalisation can be chosen to apply more power to a weaker channel realisation. 
This power inversion (PI) approach is centralised and has similarity with channel inversion \cite{haustein_performance_2002}, but avoids a matrix inversion operation:
\begin{equation}
    c_l^\text{PI} = \frac{1}{\sqrt{M}} \sum_{m=1}^M \sum_{n=1}^N |h_{ml}[n]|^2.
\end{equation}
PI is prohibitive for single antenna systems, because it could lead to an extreme peak to average power (PAP) on the antenna element.
Nevertheless, it will be shown that finite large scale antenna systems can provide enough diversity to reduce the PAP to a viable amount.

The coefficients are following $c_l^\text{PI} \geq c_l^\text{TR} \geq c_l^\text{DTR}$, if the channel realisation is weaker than the expectation $\sum_{m=1}^M \sum_{n=1}^N |h_{ml}[n]|^2 < \sqrt{M}$.
Hence, PI is inverting the behaviour of TR and DTR where less power is transmitted if the channel realisation is weak.

\section{Maximum Diversity Channel}
A bound for channel hardeningis found using an ideal maximum diversity channel.
The corresponding power delay profile (PDP) is modelled with a rectangular shape and independent identically distributed Rayleigh taps, since maximum diversity is achieved for a diffuse scattering environment if all diversity branches behave the same.
The coefficients are therefore following a zero mean circular symmetric complex normal (CN) distribution and we set the variance to $1/N$ for a $N$ tap channel to adhere to the assumption in Eqn. \eqref{eqn:pdp_normalisation}.

For DTR, $P_{mk}^\text{Ant,R}$ is a scaled sum of squares of $h_{mk}[n]$ and each squared channel coefficent follows an exponential distribution.
The scaling compensates for $c_l^\text{DTR}$ such that the result is distributed according to a Gamma distribution with shape $N$ and scale $1/N$ ($\Gamma(N,1/N)$), since the $N$ addends are independent identically distributed random variables of Gamma type ($\Gamma(1, 1/N)$) \cite{papoulis_probability_1991,mathai_storage_1982}.

The distributions for the other two normalisations diverge from $\Gamma(N,1/N)$ since each realisation of the the normalisation coefficient varies from $\sqrt{M}$.
The variance of TR will be smaller than $1/N$ since less power is applied for weaker channels.
The opposite is true for DTR because more power is applied for weaker channels.

For DTR and $M$ independent realisations of the channel coefficents over $N$ delay taps follows $P_{k}^\text{BS,R}$ $\Gamma(MN,1/(MN))$.
TR leads to a constant of one and PI has a higher variance of the relative BS transmit power due to the uncertainty of $c_l^\text{PI}$ around $\sqrt{M}$.

The remaining small scale fading for a user is captured by the variation of $P_{k}^\text{RX,R}$.
PI enforces a value of one, whilst TR leads to a distribution by $\Gamma(MN,1/(MN))$.
For DTR the result follows the square of $\Gamma(MN,1/(MN))$ being a generalised Gamma distribution \cite{stacy_generalization_1962}.
A summary of expectations and variances for the different powers is given in Tab. \ref{tab:powers_e_v} showing the scaling properties with respect to the number of taps $N$ and the number of BS antennas $M$.

\begin{table*}[]
    \centering
    \caption{Summery of expectations ($\mathcal{E}$) and variances ($\mathcal{V}$) for the relative antenna transmit power $P_{mk}^\text{Ant,R}$, the relative base station transmit power $P_{k}^\text{BS,R}$ and the relative effective received power $P_{k}^\text{RX,R}$ for a maximum diversity channel with $N$ tap normalised rectangular power delay profile for a $M$ antenna base station.
    The values are given for the different time reversal normalisations.}
    \label{tab:powers_e_v}
    \begin{tabular}{cccccccccc}
        \toprule
         & \multicolumn{3}{c}{$P_{mk}^\text{Ant,R}$} & \multicolumn{3}{c}{$P_{k}^\text{BS,R}$} & \multicolumn{3}{c}{$P_{k}^\text{RX,R}$} \\
        \cmidrule(lr){2-4} \cmidrule(lr){5-7} \cmidrule(lr){8-10}
         & DTR & TR & PI
         & DTR & TR & PI
         & DTR & TR & PI \\
        \midrule
        $\mathcal{E}$ & 1 & 1 & $\ge 1$ & 1 & 1 & $\ge 1$ & $1 + \frac{1}{MN}$ & 1 & 1\\
        $\mathcal{V}$ & $\frac{1}{N}$ & $\leq \frac{1}{N}$ & $\geq \frac{1}{N}$ & $\frac{1}{MN}$ & 0 & $\geq \frac{1}{MN}$ & $\frac{4}{MN} + \frac{10}{M^2 N^2} + \frac{6}{M N^3}$ & $\frac{1}{MN}$ & 0 \\
        \bottomrule
    \end{tabular}
\end{table*}

Naturally, the power fluctuations per antenna element are only dependent on the length of the channel, whereas the relative BS transmit and the relative received power depend on the number the BS antennas providing diversity.
We want to point the duality between BS antenna elements and channel taps in the effective channel out.
Even for non-ideal channels, both can provide diversity to compensate for small scale fading.

\section{Simulation}
To validate the derived distributions and to demonstrate the impact of the different normalisations on the relative powers simulations were conducted.
Realisations of $h_{mk}[n]$ were drawn from $\text{CN}(0, 1/N)$ to apply post processing according to Eqns. \eqref{eqn:p_ant}, \eqref{eqn:p_bs} and \eqref{eqn:p_rx}.
This allows the generation of empirical cumulative density functions (CDFs) and empirical complementary cumulative density functions (CCDFs) to simulate the behaviour of the maximum diversity channel for different finite large scale antenna system sizes.

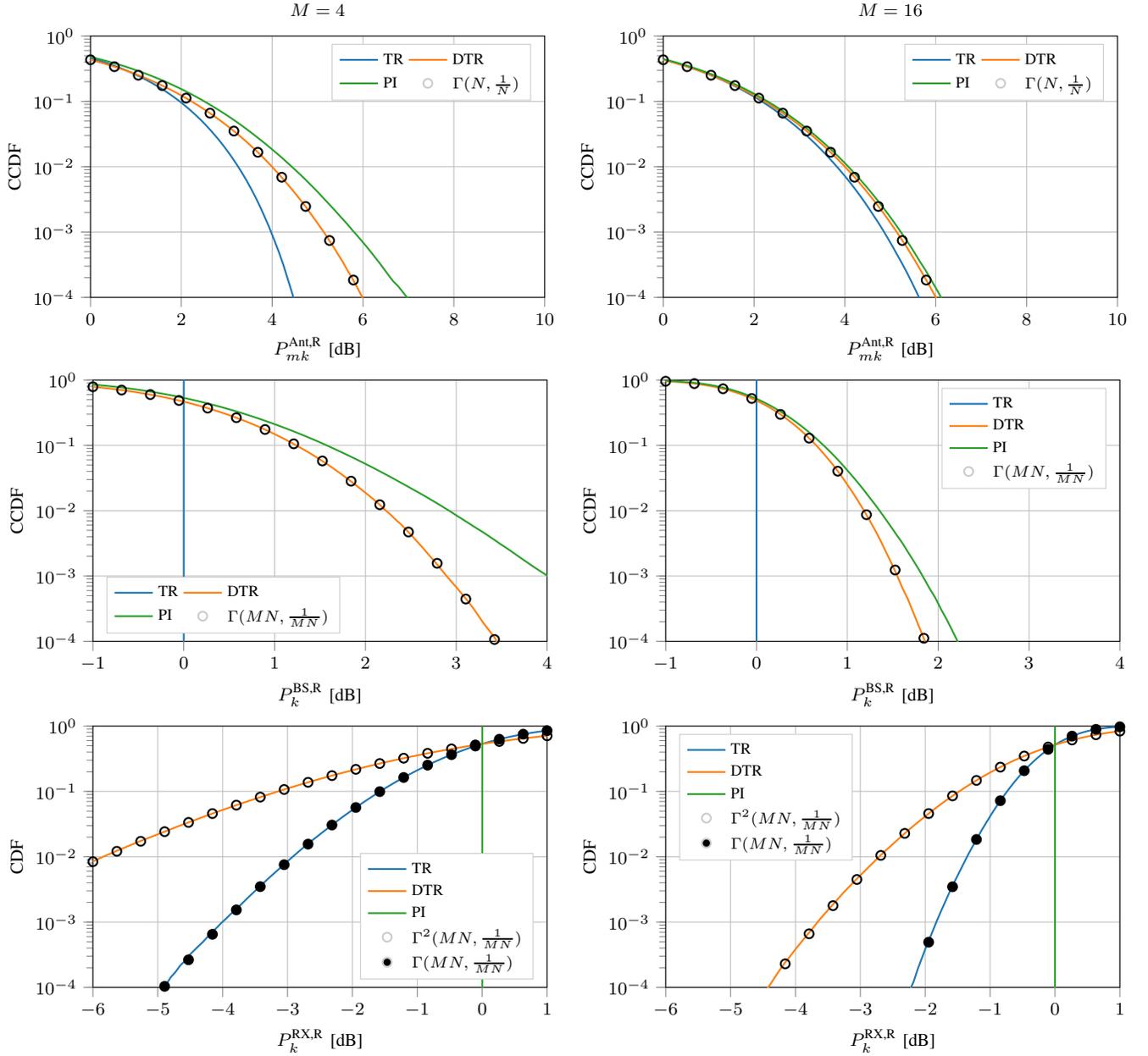
\begin{figure*}
\footnotesize
\centering
\begin{subfigure}[b]{.49\textwidth}
\centering
\begin{tikzpicture}
\begin{axis}[
title={$M=4$},
legend cell align={left},
legend entries={{TR},{DTR},{PI},{$\Gamma(N,\frac{1}{N})$}},
legend style={at={(0.97,0.97)}, anchor=north east, draw=white!80.0!black, legend columns=2,font=\scriptsize},
tick align=outside,
tick pos=left,
xlabel={$P_{mk}^\text{Ant,R}$ [dB]},
xmajorgrids,
xminorgrids,
xmin=0, xmax=10,
ylabel={CCDF},
ymajorgrids,
ymin=1e-04, ymax=1,
ymode=log,
width=.98\textwidth,height=.64\textwidth
]
\addplot [thick, color0] table {2019_sitb_p_ant_m4_n4_1.csv};
\addplot [thick, color1] table {2019_sitb_p_ant_m4_n4_2.csv};
\addplot [thick, color2] table {2019_sitb_p_ant_m4_n4_3.csv};
\addplot [thick, mark=o, only marks] table {2019_sitb_p_ant_m4_n4_4.csv};
\end{axis}
\end{tikzpicture}
\end{subfigure}
\begin{subfigure}[b]{.49\textwidth}
\centering
\begin{tikzpicture}
\begin{axis}[
title={$M=16$},
legend cell align={left},
legend entries={{TR},{DTR},{PI},{$\Gamma(N,\frac{1}{N})$}},
legend style={at={(0.97,0.97)}, anchor=north east, draw=white!80.0!black, legend columns=2,font=\scriptsize},
tick align=outside,
tick pos=left,
xlabel={$P_{mk}^\text{Ant,R}$ [dB]},
xmajorgrids,
xminorgrids,
xmin=0, xmax=10,
ylabel={CCDF},
ymajorgrids,
ymin=1e-04, ymax=1,
ymode=log,
width=.98\textwidth,height=.64\textwidth
]
\addplot [thick, color0] table {2019_sitb_p_ant_m16_n4_1.csv};
\addplot [thick, color1] table {2019_sitb_p_ant_m16_n4_2.csv};
\addplot [thick, color2] table {2019_sitb_p_ant_m16_n4_3.csv};
\addplot [thick, mark=o, only marks] table {2019_sitb_p_ant_m16_n4_4.csv};
\end{axis}
\end{tikzpicture}
\end{subfigure}
\begin{subfigure}[b]{.49\textwidth}
\centering
\begin{tikzpicture}
\begin{axis}[
legend cell align={left},
legend entries={{TR},{DTR},{PI},{$\Gamma(MN,\frac{1}{MN})$}},
legend style={at={(0.03,0.03)}, anchor=south west, draw=white!80.0!black,legend columns=2,font=\scriptsize},
tick align=outside,
tick pos=left,
xlabel={$P_{k}^\text{BS,R}$ [dB]},
xmajorgrids,
xminorgrids,
xmin=-1, xmax=4,
ylabel={CCDF},
ymajorgrids,
ymin=1e-04, ymax=1,
ymode=log,
width=.98\textwidth,height=.64\textwidth
]
\addplot [thick, color0] coordinates {
	(0,1e-6)
	(0,1)
};
\addplot [thick, color1] table {2019_sitb_p_bs_m4_n4_2.csv};
\addplot [thick, color2] table {2019_sitb_p_bs_m4_n4_3.csv};
\addplot [thick, mark=o, only marks] table {2019_sitb_p_bs_m4_n4_4.csv};
\end{axis}
\end{tikzpicture}    
\end{subfigure}
\begin{subfigure}[b]{.49\textwidth}
\centering
\begin{tikzpicture}
\begin{axis}[
legend cell align={left},
legend entries={{TR},{DTR},{PI},{$\Gamma(MN,\frac{1}{MN})$}},
legend style={at={(0.97,0.97)}, anchor=north east, draw=white!80.0!black,font=\scriptsize},
tick align=outside,
tick pos=left,
xlabel={$P_{k}^\text{BS,R}$ [dB]},
xmajorgrids,
xminorgrids,
xmin=-1, xmax=4,
ylabel={CCDF},
ymajorgrids,
ymin=1e-04, ymax=1,
ymode=log,
width=.98\textwidth,height=.64\textwidth
]
\addplot [thick, color0] coordinates {
	(0,1e-6)
	(0,1)
};
\addplot [thick, color1] table {2019_sitb_p_bs_m16_n4_2.csv};
\addplot [thick, color2] table {2019_sitb_p_bs_m16_n4_3.csv};
\addplot [thick, mark=o, only marks] table {2019_sitb_p_bs_m16_n4_4.csv};
\end{axis}
\end{tikzpicture}
\end{subfigure}
\begin{subfigure}[b]{.49\textwidth}
\centering
\begin{tikzpicture}
\begin{axis}[
legend cell align={left},
legend entries={{TR},{DTR},{PI},{$\Gamma^2(MN,\frac{1}{MN})$},{$\Gamma(MN,\frac{1}{MN})$}},
legend style={at={(0.97,0.03)}, anchor=south east, draw=white!80.0!black,font=\scriptsize},
tick align=outside,
tick pos=left,
xlabel={$P_{k}^\text{RX,R}$ [dB]},
xmajorgrids,
xminorgrids,
xmin=-6, xmax=1,
ylabel={CDF},
ymajorgrids,
ymin=1e-04, ymax=1,
ymode=log,
width=.98\textwidth,height=.64\textwidth
]
\addplot [thick, color0] table {2019_sitb_p_eff_h_m4_n4_1.csv};
\addplot [thick, color1] table {2019_sitb_p_eff_h_m4_n4_2.csv};
\addplot [thick, color2] coordinates {
	(0,1e-6)
	(0,1)
};
\addplot [thick, mark=o, only marks] table {2019_sitb_p_eff_h_m4_n4_4.csv};
\addplot [thick, mark=*, only marks] table {2019_sitb_p_eff_h_m4_n4_5.csv};
\end{axis}
\end{tikzpicture}
\end{subfigure}
\begin{subfigure}[b]{.49\textwidth}
\centering
\begin{tikzpicture}
\begin{axis}[
legend cell align={left},
legend entries={{TR},{DTR},{PI},{$\Gamma^2(MN,\frac{1}{MN})$},{$\Gamma(MN,\frac{1}{MN})$}},
legend style={at={(0.03,0.97)}, anchor=north west, draw=white!80.0!black,font=\scriptsize},
tick align=outside,
tick pos=left,
xlabel={$P_{k}^\text{RX,R}$ [dB]},
xmajorgrids,
xminorgrids,
xmin=-6, xmax=1,
ylabel={CDF},
ymajorgrids,
ymin=1e-04, ymax=1,
ymode=log,
width=.98\textwidth,height=.64\textwidth
]
\addplot [thick, color0] table {2019_sitb_p_eff_h_m16_n4_1.csv};
\addplot [thick, color1] table {2019_sitb_p_eff_h_m16_n4_2.csv};
\addplot [thick, color2] coordinates {
	(0,1e-6)
	(0,1)
};
\addplot [thick, mark=o, only marks] table {2019_sitb_p_eff_h_m16_n4_4.csv};
\addplot [thick, mark=*, only marks] table {2019_sitb_p_eff_h_m16_n4_5.csv};
\end{axis}
\end{tikzpicture}
\end{subfigure}
\caption{Empirical complementary cumulative distribution functions (CCDFs) are shown for the single antennas and the base station to highlight the excessive relative output power probabilities of a maximum diversity channel with a four tap rectangular power delay profile.
The bottom row shows the empirical cumulative distribution functions (CDFs) for the relative effective received power showing the remaining effects of small scale fading on the effective channel.
Different normalisation coefficients are used for the time reversal weights: time reversal (TR), distributed time reversal (DTR) and power inversion (PI).}
\label{fig:eccdfs_ecdfs}
\end{figure*}

An ensemble of one million realisations with $N=4$ is used for demonstration purposes and the results for $M = 4$ and $M = 16$ are presented in Fig. \ref{fig:eccdfs_ecdfs}.
The CCDFs of the first row show the distribution of relative antenna element output power due to the realisations of the channel coefficients.
As expected, TR and PI require less and more excess power (with respect to a reference at \SI{0}{\deci\bel}) then DTR, respectively.
The differences to DTR are vanishing for growing $M$, since $c_l^\text{TR}$ and $c_l^\text{PI}$ are converging to $c_l^{DTR} = \sqrt{M}$ due to the self-averaging properties of the large scale antenna system.

The CCDF of $P_{k}^\text{Ant,R}$ can for moderately sized finite large scale antenna systems be predicted from the distribution of the channel coefficients.
For $M=16$ do the requirements for PI and MRT vary less then \SI{0.5}{\deci\bel} with respect to DTR.

The next row in Fig. \ref{fig:eccdfs_ecdfs} shows the CCDFs for the relative BS transmit power.
Both, DTR and PI require more excess sum power then TR, but the difference is reduced for larger $M$.
It is important to note that the reduction comes from averaging over multiple realisations of antenna output powers.
Hence, the unit normalisation of TR is giving no insight into how the antenna element output powers are behaving.
It only ensures that the sum over all antenna elements becomes one for the specific user.

The bottom row in Fig. \ref{fig:eccdfs_ecdfs} shows the CDFs for the relative received power at a user.
The distributions describe the remaining small scale fading directly.
PI compensates it completely whilst TR and DTR reduce it's severity.
DTR is prone to a doubling in \si{\deci\bel} with respect to TR but opens up for distributed weight calculations.
The trade-off between distributed and centralised weight calculation is directly accessible from the distribution of $P_{k}^\text{RX,R}$.
Furthermore, the results show how much the channel hardening is exploited by the different time reversal normalisations.

\begin{figure*}[]
    \footnotesize
    \centering
    \begin{tikzpicture}
\begin{axis}[
title={$N = 4$},
legend cell align={left},
legend entries={{TR},{DTR},{PI}},
legend style={at={(0.97,0.03)}, anchor=south east, draw=white!80.0!black, legend columns=3,font=\scriptsize},
tick align=outside,
tick pos=left,
xlabel={$M$ [base station antennas]},
xmajorgrids,
xminorgrids,
xmin=1, xmax=128, xmode=log,
ylabel={$P^\text{\{Ant,BS,RX\},R}$ [dB]},
ymajorgrids,
ymin=-10, ymax=10,
width=0.98\textwidth,height=.34\textwidth
]
\addplot [thick, color0] table[x=M, y=P_ant_MRT] {data_points.csv};
\addplot [thick, color1] table[x=M, y=P_ant_sqrt] {data_points.csv};
\addplot [thick, color2] table[x=M, y=P_ant_PI] {data_points.csv};
\draw (axis cs:20,6) node[draw, ellipse, minimum height=0.5cm,minimum width=0.25cm] (P_ant) {} node[above, yshift=0.25cm] {$P_{mk}^\text{Ant,R}$};
\addplot [thick, color0,dashed] table[x=M, y=P_bs_MRT] {data_points.csv};
\addplot [thick, color1, dashed] table[x=M, y=P_bs_sqrt] {data_points.csv};
\addplot [thick, color2, dashed] table[x=M, y=P_bs_PI] {data_points.csv};
\draw (axis cs:30,1) node[draw, ellipse, minimum height=0.7cm,minimum width=0.25cm] (P_ant) {} node[above, yshift=0.35cm] {$P_{k}^\text{BS,R}$};
\addplot [thick, color0, dotted] table[x=M, y=h_e_MRT] {data_points.csv};
\addplot [thick, color1, dotted] table[x=M, y=h_e_sqrt] {data_points.csv};
\addplot [thick, color2, dotted] table[x=M, y=h_e_PI] {data_points.csv};
\draw (axis cs:50,-1.3) node[draw, ellipse, minimum height=0.7cm,minimum width=0.25cm] (P_ant) {} node[below, yshift=-0.4cm] {$P_{k}^\text{RX,R}$};
\end{axis}
\end{tikzpicture}
    \caption{
    For the same simulation scenario as in Fig. \ref{fig:eccdfs_ecdfs} are relative power levels displayed. $P_{mk}^\text{Ant,R}$ and $P_{k}^\text{BS,R}$ represent the antenna element and the whole base station, respectively. 
    The displayed values are exceeded with a probability of $10^{-4}$. 
    Additionally, relative power levels at the receiver $P_{k}^\text{RX,R}$ fall short of the shown value with the same probability.}
    \label{fig:m_scaling}
\end{figure*}
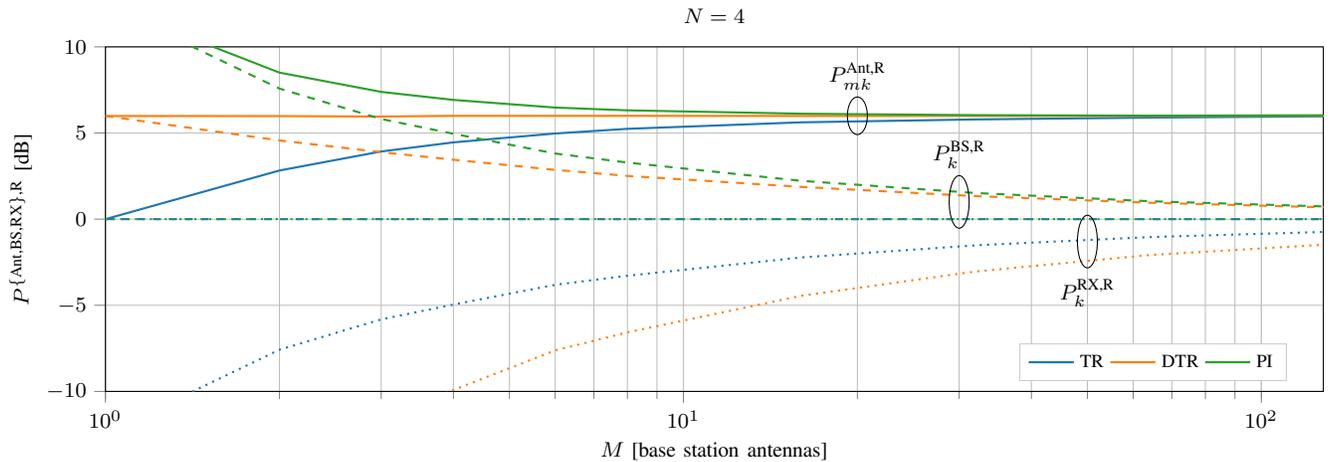

Fig. \ref{fig:m_scaling} shows how the empirical CCDFs and CDFs behave at a probability of $10^{-4}$ for growing $M$ and $N=4$.
Channel hardening leads to fast convergence of the relative antenna element power to \SI{6}{\deci\bel}, no matter the chosen normalisation.
In addition, the relative power of the whole BS is converging towards the TR constant of \SI{0}{\deci\bel}.
The penalty of excess power between PI and DTR is vanishing around 32 antenna elements.
Finally, small scale fading has a diminishing effect for TR and DTR.
In summary, the figure shows the trade-offs for a four tap maximum diversity channel.
If small scale fading is supposed to be mitigated completely, then PI could be used if a slight excess in output power from each antenna is acceptable.
Each transmitter for a 32 antenna system would have to supply about \SI{0.2}{\deci\bel} more excess power then TR, leading to an increased BS output power of \SI{1.5}{\deci\bel} in less then $10^{-4}$ cases.

The simulated findings encourage to incorporate realistic PDPs for comparison to the ideal maximum diversity channel.
Additionally, measurements could provide the realisations for the empirical CCDFs and CDFs in realistic environments.
Ultimately, fully synchronised uplink and downlink measurements should be conducted to verify that PI can completely compensate for small scale fading without exceeding a certain PAP requirement.

\section{Conclusion}

This paper considers the effective massive MIMO channel in the time domain to analyse the severity of small scale fading for an ideal maximum diversity channel.
This approach bounds the remaining small scale fading and shows the exploitation of channel hardening.
Time reversal precoding with different normalisations is described and the impact on relative transmitter power, sum BS power and effective received power for a single tap receiver is demonstrated.
Furthermore, distributions for the the relative powers with DTR normalisation are given.
They can be used to bound the remaining small scale fading for system design purposes.

For large scale antenna systems, the actual normalisation coefficient has little impact on the relative excess transmit power requirement for each BS antenna element, but influences the excess sum BS power.
The latter is merely of regulatory interest and depends on the averaging time window given by the authorities, since each single transmitter needs to fulfil it's PAP requirements nonetheless.

A time reversal precoder can allow for either distributed (DTR) or centralised (TR and PI) weight calculations.
DTR relaxes the requirements on inter BS communication, since all fast weight calculations can be done locally at each antenna element.
However, additional power needs to be spent to guarantee a specified downlink performance as the remaining small scale fading is larger then for TR.

If the system design allows for centralised weight calculation, then PI can be chosen over TR to compensate for the remaining small scale fading.
The penalty is a slightly fluctuating relative BS power to realise a fixed relative received power at the user, whilst the requirements for the relative transmitter power increases negligibly.

The present study suggests that PI is realisable for environments with sufficient spatial diversity.
An ideal 16 antenna system observing a maximum diversity four tap channel provides 64 degrees of freedom and the penalty for increasing the robustness of the link is as small as \SI{0.5}{\deci\bel} excess power per antenna element in $10^{-4}$ cases.
The overall BS power has an expectation of around one and exceeds it in less then $10^{-4}$ cases by \SI{2.2}{\deci\bel}.
The BS excess power is mainly of regulatory interest because the BS has to provide similar transmitters for all presented normalisations.
Eventually, the resulting effective downlink channel can compensate for small scale fading, leaving the system engineer to consider large scale fading for the design of WSNs.

\bibliography{2019_sitb}
\bibliographystyle{ieeetr}

\end{document}